\documentclass[3p]{elsarticle}
\UseRawInputEncoding
\usepackage{lineno,hyperref}
\usepackage{graphicx}
\usepackage{amsmath}
\usepackage{color}
\usepackage{float}
\usepackage[normalem]{ulem}
\usepackage{mathtools}
\usepackage{rotating}
\usepackage{lscape}
\usepackage{booktabs}
\usepackage{subcaption}
\usepackage{placeins}
\usepackage{array}
\usepackage{tabularx}

\journal{arXiv}
\begin{document}

\begin{frontmatter}

\title{Fitting Accumulated Stock Returns with Tempered Skew t-Distribution} 

\author[mymainaddress]{Siqi Shao}
\author[mymainaddress]{R. A. Serota\fnref{myfootnote}}
\fntext[myfootnote]{serota@ucmail.uc.edu}

\address[mymainaddress]{Department of Physics, University of Cincinnati, Cincinnati, Ohio 45221-0011}

\begin{abstract}
We analyze distributions of historic S\&P500 multi-day returns, for the number of days of accumulation from 20 to 120. With the increase of the number of days of accumulation, we observe clear tempering of power-law tails toward a seemingly finite value. To explain this phenomenon, we employ a model that produces a "capped Inverse Gamma" stationary (steady-state) distribution for stochastic volatility which, in turn, produces a "tempered Student-t" distribution for returns. We then employ Jones-Faddy-like symmetry breaking mechanism that produces a "tempered Skew-t" distribution. This distribution provides rather good fits to the distributions of accumulated multi-day S\&P500 returns, which exhibit symmetry breaking between gains and losses -- as reflected by positive mean and negative skew. Tempered Skew-t fits are also consistent with near perfect linear dependence on the number of days of accumulation of the mean values and, even more so, of the variances (mean squared realized volatility) of the distributions. 
\end{abstract}


\end{frontmatter}

\section{Introduction}

Student's t-distribution has a long history in statistics and finds multiple applications \cite{fisher1925applications,martin1971statistics,ahsanullah2014normal}. In its simplest form its probability density function (PDF) is given by
\begin{equation}
f_{\rm St}(x)=\frac{\Gamma\left(\frac{\nu+1}{2}\right)}{(\pi \nu)^{\frac{1}{2}}\Gamma\left(\frac{\nu}{2}\right)} \left(1+\frac{x^2}{\nu} \right)^{-\frac{\nu+1}{2}}=\frac{1}{\sqrt{\nu} B\left(\frac{\nu}{2}, \frac{1}{2}\right)} \left(1+\frac{x^2}{\nu} \right)^{-\frac{\nu+1}{2}}
\label{fst}
\end{equation}
where $\Gamma$ and $B$ are Gamma and Beta functions respectively \cite{nist2025digital} and $\nu$ is the shape parameter referred to as "degrees of freedom``. Introduction of a scale parameter $\sigma$ and location parameter $\mu$ \cite{wolfram2025student} yields a simple generalization of (\ref{fst}) 
\begin{equation}
f_{\rm Stg}(x)=\frac{1}{\sqrt{\nu} \sigma B\left(\frac{\nu}{2}, \frac{1}{2}\right)} \left(1+\frac{(x-\mu)^2}{\nu \sigma^2} \right)^{-\frac{\nu+1}{2}}
\label{fstg}
\end{equation}
Using an identity
\begin{equation}
\left(1+\frac{x^2}{\nu} \right)^{-1}=\left(1+\frac{x}{\sqrt{x^2+\nu}}\right)\left(1-\frac{x}{\sqrt{x^2+\nu}}\right)
\label{identity}
\end{equation}
and giving different powers to the left and right parentheses, $\nu/2 \rightarrow \nu_l/2$, and $\nu/2 \rightarrow \nu_r/2$,  one can construct from (\ref{fst}) a Jones-Faddy (JF) \cite{jones2001skew,jones2003skew} Skew t-distribution
\begin{equation}
f_{\rm JF}(x)= \frac{1}{2^{\bar{\nu} -1} \sqrt{\bar{\nu}}B\left(\frac{\nu_l}{2}, \frac{\nu_r}{2} \right)} \left(1+\frac{x}{\sqrt{x^2+\bar{\nu}}}\right)^{\frac{\nu_l+1}{2}} \left(1-\frac{x}{\sqrt{x^2+\bar{\nu}}}\right)^{\frac{\nu_r+1}{2}}, \qquad \bar{\nu} = \frac{\nu_l + \nu_r}{2}
\label{fJF}
\end{equation}
An important feature of the distribution (\ref{fJF}) is that the exponents of the power-law tails at infinities are different:
\[
f_{\rm JF}(x) \propto |x|^{-(\nu_l+1)}, \qquad x \rightarrow - \infty
\]
and
\[
f_{\rm JF}(x) \propto x^{-(\nu_r+1)}, \qquad x \rightarrow \infty
\]
Another important feature of JF Skew-t distribution is that it can be obtained from Beta distribution by a change of variable, which allows to easily derive its moments and its cumulative distribution function (CDF) \cite{jones2001skew,jones2003skew}.

Student's t-distribution also emerges in analysis of stock market returns \cite{praetz1972distribution,fuentes2009universal,liu2019distributions} from a pair of stochastic differential equations (SDE). Namely, a mean-reverting multiplicative model for stochastic variance \cite{nelson1990arch}
\begin{equation}
\mathrm{d}v_t = -\gamma(v_t - \theta)\mathrm{d}t + \kappa v_t \mathrm{d}W^{(2)}
\label{dvt}
\end{equation}
which yields an Inverse Gamma \cite{wolfram2025inverse} PDF of the stationary distribution of $v_t $ \cite{praetz1972distribution,fuentes2009universal,liu2019distributions}, coupled with the stochastic model for the fluctuations of stock returns relative to the overall gains
\begin{equation}
dx_t \equiv  \log\left(\frac{S_{t+\mathrm{d}t}}{S_t}\right) - \mu \mathrm{d}t = \sigma_t \mathrm{d}W^{(1)}
\label{dxt}
\end{equation}
yields, via a product distribution of $\sigma_t$ and $\mathrm{d}W^{(1)}$ \cite{ma2014model}, a Student-t distribution of stock returns 
\begin{equation}
f_{\rm Str}(x)=
\frac{\Gamma(\frac{\alpha}{\theta} + \frac{3}{2})}{\sqrt{\pi}\Gamma(\frac{\alpha}{\theta}+1)}\frac{1}{\sqrt{2\alpha \tau}}\left( \frac{x^2}{2\alpha \tau} + 1\right)^{-(\frac{\alpha}{\theta} + \frac{3}{2})}, \qquad \alpha= \frac{2 \gamma \theta}{\kappa^2}
\label{fStr}
\end{equation}
Here $\sigma_t$ is the stochastic volatility, $v_t=\sigma_t^2$ is the stochastic variance, $\tau$ is the number of days of accumulation of returns, $\mathrm{d}W = W\left(t+\mathrm{d}t\right) - W(t)$ is the normally distributed Wiener process, $\mathrm{d}W \sim \mathrm{N(}0,\, \mathrm{d}t \mathrm{)}$, $(\mathrm{d}W)^2 = \mathrm{d}t$, $S_t$ is the security (stock) price on day $t$, and $\mu$ is the slope of the best linear fit of $\log\left( S_T/S_0\right)$, T being the period over which returns are studied -- 1980 to 2025 S\&P500 returns here.

Student-t distribution (\ref{fStr}) implies symmetry between gains and losses, even with the introduction of the location parameter $\mu : x \rightarrow x-\mu$. This symmetry, however, is broken in empirical data which exhibits negative skew, positive mean and heavier tails of losses versus gains \cite{farahani2025asymmetry}. In order to describe these effects we introduced a modified Jones-Faddy (mJF) Skew-t distribution for returns \cite{shao2026broken} whose PDF is given by
\begin{equation}
f_{\rm mJFr}(x)=C\left(1-\frac{x-\mu}{\sqrt{(x-\mu)^2+(\alpha_g+\alpha_l) \tau}}\right)^{\frac{\alpha_g}{\theta}+\frac{3}{2}} \left(1+\frac{x-\mu}{\sqrt{(x-\mu)^2+(\alpha_g+\alpha_l) \tau}}\right)^{\frac{\alpha_l}{\theta}+\frac{3}{2}}
\label{fmJFr}
\end{equation}
where $C$ is the normalization constant \cite{shao2026broken}, and which accounts rather well for aforementioned features of empirical distributions \cite{ghasemi2026broken} -- at least for relatively small numbers of days of accumulation $\tau=1,2,\ldots,10$. Nonetheless, while (\ref{fmJFr}) predicts power-law tails 
\[
 f_{\rm mJFr} \propto |x|^{-\left( 2 \frac{\alpha_l}{\theta}+ 3 \right)}, \qquad x \rightarrow -\infty
\]
and 
\[
f_{\rm mJFr} \propto x^{-\left( 2 \frac{\alpha_g}{\theta}+ 3 \right)}, \qquad x\rightarrow \infty
\]
for losses and gains respectively, the empirical distributions start exhibiting distinct tempering of the tail ends already on $\tau$ approaching $10$  \cite{ghasemi2026broken}. This tempering becomes substantially more pronounced for larger $\tau$ where power-law dependence gives way to a decay to seemingly finite values.

The purpose of this work, therefore, is to try to explain transition from pure power-law dependence to tempering and possible termination of tails. Towards this end we introduce a novel SDE for stochastic variance which yields a "capped Inverse Gamma" as its stationary distribution. This, in turn, produces a "tempered Student t-distribution" which is subsequently transformed via symmetry breaking into a "tempered mJF Skew-t" distribution for returns. Analytical derivations are given in Section \ref{analytic} and their predictions are tested numerically vis-a-vis empirical data in Section \ref{numerics}. Section \ref{scaling} addresses scaling characteristics of the distributions. We conclude with the discussion of our results and possible directions of future work.

\section{Analytic Derivations\label{analytic}}
\subsection{Capped Inverse Gamma Distribution \label{tIGa}}
Capped Inverse Gamma Distribution is a stationary state distribution of stochastic volatility obtained from a generalized version of SDE (\ref{dvt}) 
\begin{equation}
\mathrm{d}v_t = -\gamma(v_t - \theta)\mathrm{d}t + \kappa_2 v_t \sqrt{1-\kappa_1^2 v_t} \mathrm{d}W^{(2)}
\label{tdvt}
\end{equation}
via Fokker-Planck formalism \cite{risken1996fokker,jacobs2010stochastic} and its PDF is given by
\begin{equation}
 \mathrm{IGa1}(\nu_t;\beta,\alpha, \kappa_1)=
 \mathcal N\,
 \nu_t^{-\beta-1}
 \left(1-\kappa_1^2\nu_t\right)^{\beta-2}
 \exp\left(-\frac{\alpha}{\nu_t}\right),
 \label{tfvt}
\end{equation}
where
\begin{equation}
 \alpha=\frac{2\gamma\theta}{\kappa_2^2},
 \qquad
 \beta=1+\frac{\alpha}{\theta}\left(1-\theta\kappa_1^2\right),
 \qquad
 \mathcal N =
 \frac{\theta e^{\alpha\kappa_1^2}\alpha^{\beta-1}}
 {\Gamma(\beta-1)}
 \label{tfvtparams}
\end{equation}
Notation ``IGa1" in (\ref{tfvt}) is chosen to reflect distributions with a finite cutoff as in, for instance, \cite{mcdonald1995generalization}. Obviously, it is implied that $ 0<\nu_t<\kappa_1^{-2}$ and, since mean value of stochastic variance $<v_t>=\theta$ in the stationary state, it also follows that $0<\theta<\kappa_1^{-2}$. For $\kappa_1=0$, (\ref{tfvt}) trivially reduces to Inverse Gamma $\mathrm{IGa}\left(\nu_t;\frac{\alpha}{\theta}+1,\alpha\right)$, which is the stationary state PDF of (\ref{dvt}) \cite{liu2019distributions}. Notice also that if $\alpha \ll  \kappa_1^{-2}$ then IGa1 of (\ref{tfvt}) would exhibit standard IGa tails for $\alpha \ll v_t \ll \kappa_1^{-2}$,
\[
\mathrm{IGa1}(v_t) \approx \mathrm{IGa}(v_t) \propto v_t^{-\left(\frac{\alpha}{\theta}+2\right)}
\]
and, consequently, PDF of stochastic volatility $\sigma_t$ would scale for $\alpha^{1/2} \ll \sigma_t \ll \kappa_1^{-1}$ as \cite{liu2019distributions}
\[
2\sigma_t \mathrm{IGa1}(\sigma_t^2)  \approx 2\sigma_t \mathrm{IGa}(\sigma_t^2) \propto \sigma_t^{-\left(2\frac{\alpha}{\theta}+3\right)}
\]
Obviously, the above power-law exponent of the volatility carries over, via a product distribution of $\sigma_t$ and $\mathrm{d}W^{(1)}$ in (\ref{dxt}) \cite{ma2014model}, to the power-law exponent of returns as is seen from (\ref{fStr}).

\subsection{Tempered Student t-Distribution \label{tSt}}

Tempered Student t-Distribution for returns is obtained, per (\ref{dxt}) via product distribution \cite{rohatgi1976introduction,grimmett2020probability} \cite{ma2014model} of stochastic volatility, whose PDF is $2\sigma_t \mathrm{IGa1}\left(\sigma_t^2; \beta,\alpha, \kappa_1 \right)$, and normally distributed Wiener increments $\mathrm{d}W \sim \mathrm{N(}0,\, \mathrm{d}t \mathrm{)}$. As a result, and introducing location parameter $\mu$, we find 
\begin{equation}
 f_{\rm tStr}(x)
 = \frac{\mathcal N}{\sqrt{2\pi\tau}}\,\mathcal I(x)
 \label{ftStr}
\end{equation}
where $ \mathcal N$ is given by (\ref{tfvtparams}),
\begin{align}
\mathcal I(x)
&= \frac{3\kappa_1^{1+2\beta}\Gamma\!\left(-\tfrac12-\beta\right)\Gamma(\beta-1)}{4\sqrt\pi}
\,{}_1F_1\!\left(
\frac52;\frac32+\beta;
-\frac{\kappa_1^2((x-\mu)^2+2\alpha\tau)}{2\tau}
\right)
\notag \nonumber \\[4pt]
&\quad
+ 
\Gamma\!\left(\beta+\frac12\right)
\alpha^{-\beta-\frac12}\left(\frac{2\alpha\tau}{(x-\mu)^2+2\alpha\tau}\right)^{\beta+\frac12}
{}_1F_1\!\left(
2-\beta;\frac12-\beta;
-\frac{\kappa_1^2((x-\mu)^2+2\alpha\tau)}{2\tau}
\right).
\label{Ix}
\end{align}
and 
 \({}_1F_1(a;b;x)\) is the confluent hypergeometric (Kummer's) function of the first kind.
A quick check shows that for $\kappa_1=0$ (\ref{ftStr}) reduces to (\ref{fStr}) with location parameter $\mu$. Indeed, in this case the first term in (\ref{Ix}) vanishes and the second term simplifies to
\begin{equation}
 \mathcal I(x, \kappa_1=0)
 = 2^{\frac{\alpha}{\theta}+\frac{3}{2}}
 \Gamma\!\left(\frac{\alpha}{\theta}+\frac{3}{2}\right)
 \left(\frac{\tau}{(x-\mu)^2+2\alpha\tau}\right)^{\frac{\alpha}{\theta}+\frac{3}{2}}
 \label{I0}
\end{equation}
so that 
\begin{equation}
 f_{\rm tStr}(x, \kappa_1=0)
 = \frac{ \Gamma\!\left(\frac{\alpha}{\theta}+\frac{3}{2}\right)}{\sqrt{\pi}\Gamma(\frac{\alpha}{\theta}+1)}\frac{1}{\sqrt{2\alpha \tau}}
 \left(\frac{(x-\mu)^2}{2\alpha \tau}+1\right)^{-(\frac{\alpha}{\theta}+\frac{3}{2})}
 \label{ftStr0}
\end{equation} 
as stated above.
\pagebreak

Eq. (\ref{ftStr0}) and (\ref{Ix}) also allow us to establish the range of power-law tails. Namely, from (\ref{ftStr0}) it is clear that the onset of power-law tails occurs approximately at
\begin{equation}
\frac{(x_o-\mu)^2}{2\alpha \tau} \sim 1
\label{onset}
\end{equation}
On the other hand, (\ref{ftStr0}) is also a zero-order approximation of the expansion in $\kappa_1$  of (\ref{ftStr}) via (\ref{Ix}). Effectively, it is an expansion in the $\kappa_1$ term inside ${}_1F_1$ in (\ref{Ix}). Consequently, an estimate for the point where power-law tail begins to bend towards its final value is given by
\begin{equation}
\frac{\kappa_1^2(x_b-\mu)^2}{2 \tau} \sim 1
\label{bend}
\end{equation}

\subsection{Tempered Modified Jones-Faddy Skew-t  Distribution \label{tmJFt}}
Symmetry of the tempered Student t-Distribution can be broken, as before \cite{shao2026broken} - see identity (\ref{identity}) and mJF Skew-t distribution of returns (\ref{fmJFr}) - by replacing the factor in (\ref{Ix}) as follows:
\begin{equation}
\left(\frac{2\alpha\tau}{(x-\mu)^2+2\alpha\tau}\right)^{\beta+\frac12}\to \left(1+\frac{x-\mu}{\sqrt{(x-\mu)^2+2\alpha\tau}}\right)^{\beta_l+\frac12}
 \left(1-\frac{x-\mu}{\sqrt{(x-\mu)^2+2\alpha\tau}}\right)^{\beta_g+\frac12}
 \label{skewfactor}
\end{equation}
which yields the following tempered mJF Skew-t PDF for returns
\begin{equation}
 f_{\rm tmJFr}(x) = \frac{1}{Z} \mathcal I_{\beta_l,\beta_g}(x)
\label{ftmJFr}
\end{equation}
with
\begin{align}
\mathcal I_{\beta_l,\beta_g}(x)
&=
\frac{3\kappa_1^{1+2\bar\beta}
\Gamma\!\left(-\frac12-\bar\beta\right)\Gamma(\bar\beta-1)}
{4\sqrt{\pi}}\,
{}_1F_1\!\left(
\frac52;\frac32+\bar\beta;
-\frac{\kappa_1^2\left[(x-\mu)^2+2\alpha\tau\right]}{2\tau}
\right)
\notag\\[3pt]
&\quad
+
\Gamma\!\left(\bar\beta+\frac12\right)
\alpha^{-\bar\beta-\frac12}
\left[
1+\frac{x-\mu}{\sqrt{(x-\mu)^2+2\alpha\tau}}
\right]^{\beta_l+\frac12}
\left[
1-\frac{x-\mu}{\sqrt{(x-\mu)^2+2\alpha\tau}}
\right]^{\beta_g+\frac12}
\notag\\[3pt]
&\qquad\qquad\times
{}_1F_1\!\left(
2-\bar\beta;\frac12-\bar\beta;
-\frac{\kappa_1^2\left[(x-\mu)^2+2\alpha\tau\right]}{2\tau}
\right), \qquad \bar{\beta}=\frac{\beta_g+\beta_l}{2}
\label{Ixgl}
\end{align}
where
$Z$ is the normalization constant, such that
\begin{equation}
\int_{-\infty}^{\infty} f_{\rm tmJFr} (x) \,dx =1
\label{Z}
\end{equation}
It should be pointed out that, with the introduction of tempered mJF Skew-t distribution, $\kappa_2$, $\gamma$ and $\theta$ lose their meaning as SDE entrants. In other words, $\alpha$, $\kappa_1$ and $\beta_g$ and $\beta_l$ (as well as $\mu$) are considered independent parameters used for fitting S\&P500 data. So if we wanted to treat $\theta$ as the remaining vestige of the mean in the mean-reverting models, the second of (\ref{tfvtparams}), for instance, should be rewritten as  
\begin{equation}
 \theta=\frac{\alpha}{\bar\beta-1+\alpha\kappa_1^2},
 \label{theta}
\end{equation}
that is $\theta$ should be considered a dependent quantity. 

\section{Simulations of S\&P500 with Tempered mJF Skew-t Distribution for Returns \label{numerics}}

\subsection{Fitting Parameters \label{params}}
\begin{figure}[H]
    \centering
    \includegraphics[width=0.75\linewidth]{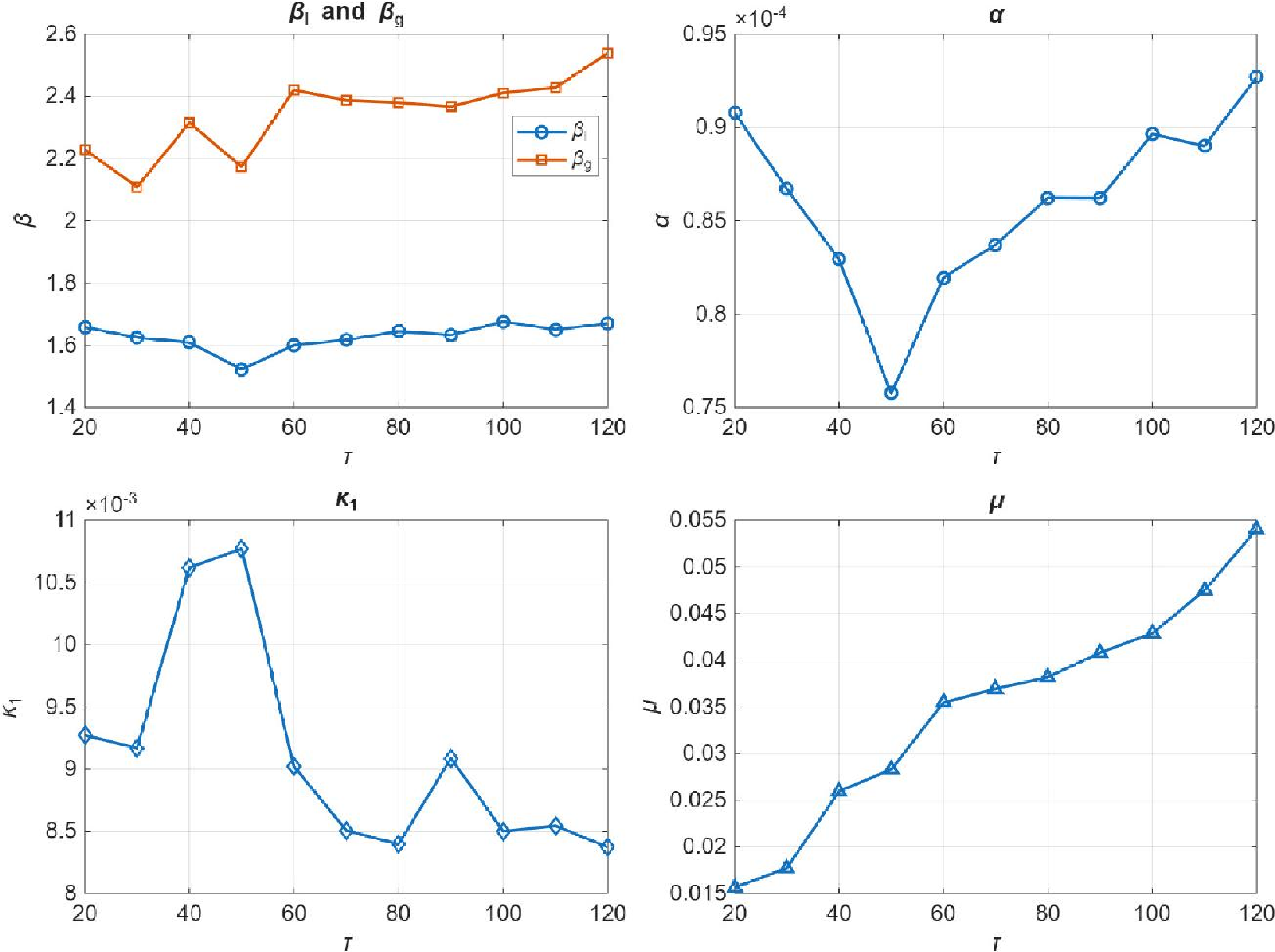}
    \caption{Fitted parameters of tempered mJF Skew-t distribution for returns as a function of number of days of accumulation $\tau$.}
    \label{fitparamsfig}
\end{figure}

Table~\ref{fitparams} lists fitted parameters $\beta_l$, $\beta_g$, $\alpha$, $\kappa_1$, and $\mu$ for S\&P500 returns accumulated over $\tau=20,30,\ldots,120$ trading days. Fig. \ref{fitparamsfig} shows its results in graphical form.
\begin{table}[H]
\centering
\begin{tabular}{@{}c c c c c c@{}}
\toprule
$\tau$ & $\beta_l$ & $\beta_g$ & $\alpha$ & $\kappa_1$ & $\mu$  \\
\midrule
20 & 1.658 & 2.229 & $9.079\times 10^{-5}$ & 0.009273 & 0.01565  \\
30 & 1.625 & 2.109 & $8.672\times 10^{-5}$ & 0.009169 & 0.01773  \\
40 & 1.609 & 2.315 & $8.293\times 10^{-5}$ & 0.010620 & 0.02596  \\
50 & 1.523 & 2.174 & $7.576\times 10^{-5}$ & 0.010770 & 0.02829  \\
60 & 1.600 & 2.420 & $8.196\times 10^{-5}$ & 0.009020 & 0.03547  \\
70 & 1.617 & 2.387 & $8.372\times 10^{-5}$ & 0.008508 & 0.03695  \\
80 & 1.645 & 2.380 & $8.623\times 10^{-5}$ & 0.008397 & 0.03818  \\
90 & 1.633 & 2.367 & $8.621\times 10^{-5}$ & 0.009087 & 0.04080  \\
100 & 1.676 & 2.411 & $8.965\times 10^{-5}$ & 0.008502 & 0.04287  \\
110 & 1.651 & 2.428 & $8.901\times 10^{-5}$ & 0.008546 & 0.04748  \\
120 & 1.670 & 2.539 & $9.270\times 10^{-5}$ & 0.008374 & 0.05405  \\
\bottomrule
\end{tabular}
\caption{Parameters obtained by fitting S\&P500 returns with tempered mJF Skew-t distribution for returns (\ref{ftmJFr}).}
\label{fitparams}
\end{table}
As was explained above, the exponents of the power-law tails of PDF between the onset of tails and their bending points are $-(2 \alpha_l/\theta+3)$ and $-(2 \alpha_g/\theta+3)$ for losses and gains respectively. Consequently, for complementary cumulative distribution functions (CCDF) they are, respectively, $-2( \alpha_l/\theta+1)$ and $-2( \alpha_g/\theta+1)$. In other words, pre-tempering CCDF slopes are approximately
\begin{equation}
 S_l\simeq -2\beta_l,
 \qquad
 S_g\simeq -2\beta_g
 \label{slopes}
\end{equation}
as seen by setting $\kappa_1=0$ in the expression (\ref{tfvtparams}) whilst $1+\alpha_{l,g}/\theta \rightarrow \beta_{l,g}$. Notice that in Fig. \ref{fitparamsfig} both $\beta$, and consequently the slopes of the pre-tempered tails, show  little variation with the number of days of accumulation of returns $\tau$ -- this is especially true for losses. Also, $\beta_l < \beta_g$ for any $\tau$, meaning that power-law parts of losses are slower than those of gains.

\begin{figure}[H]
    \centering
    \includegraphics[width=0.72\textwidth]{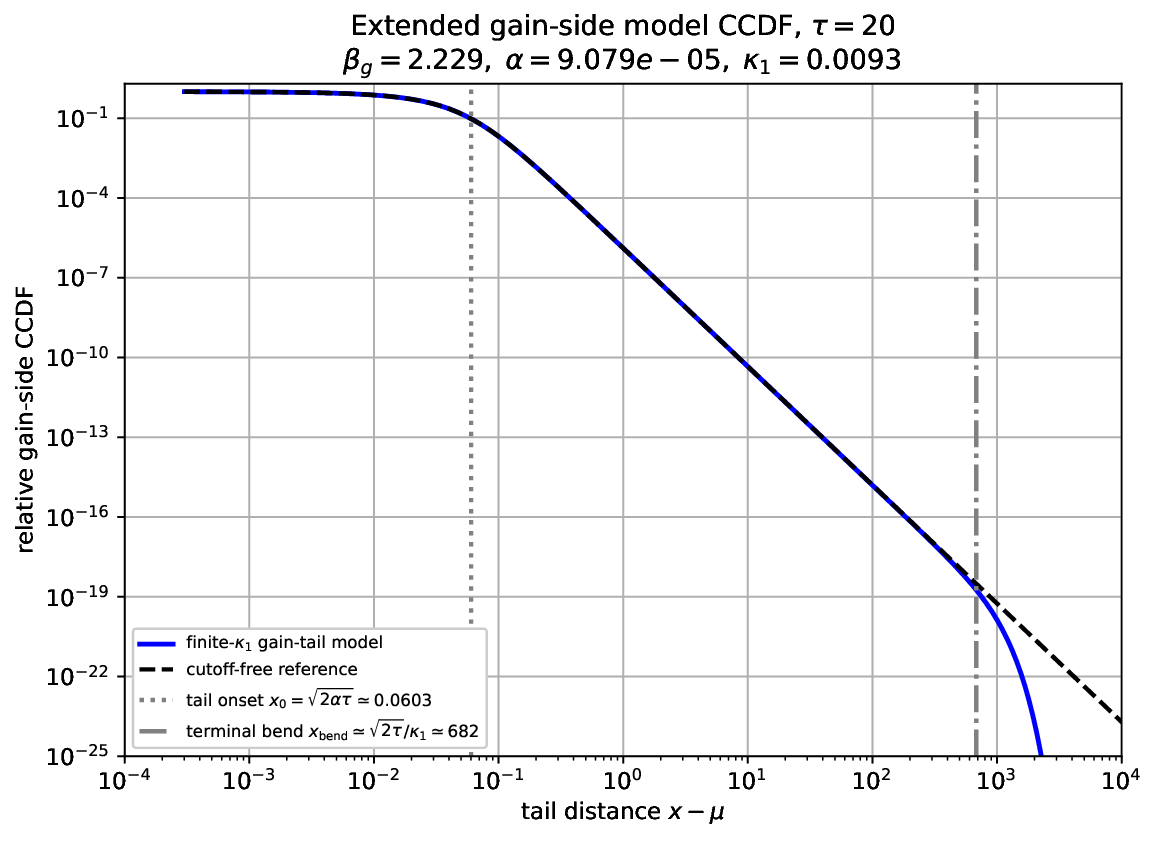}
    \caption{Gain-side CCDF from PDF (\ref{ftmJFr}) with parameters from Table \ref{fitparams} for $\tau=20$. 
    Power-law-like behavior is in effect between approximate common onset $x_o$ from (\ref{onset}) and approximate common bending scale $x_b$ from (\ref{bend}).}
    \label{tempering}
\end{figure}

Table \ref{derived} shows parameters derived from (\ref{onset}), (\ref{bend}), (\ref{theta}), and (\ref{slopes}) using fitting parameters from Table \ref{fitparams}. Fig. \ref{tempering} illustrates the onset of power-law tails and their eventual bending.
\begin{table}
\centering
\caption{Cut-off parameters for stochastic variance $\kappa_1^{-2}$ and stochastic volatility $\kappa_1^{-1}$; $\theta$ from (\ref{theta}); approximate common tail-distance onset $x_o-\mu$ and common tail-distance bending scale $x_b-\mu$ from (\ref{onset}) and (\ref{bend}); approximate pre-bending slopes of losses $S_l$ and gains $S_g$ from (\ref{slopes}). All values are obtained using fitting parameters from Table \ref{fitparams} for $\tau=20,30,\ldots,120$ days of accumulation.}
\begin{tabular}{@{}c c c c c c c c@{}}
\toprule
$\tau$ & $\kappa_1^{-2}$ & $\kappa_1^{-1}$ & $\theta$ & $x_o-\mu$ & $x_b-\mu$ & $S_l$ & $S_g$  \\
\midrule
20 & $1.163\times 10^{4}$ & 107.84 & $9.623\times 10^{-5}$ & $6.026\times 10^{-2}$ & $6.820\times 10^{2}$ & -3.316 & -4.458  \\
30 & $1.189\times 10^{4}$ & 109.06 & $1.000\times 10^{-4}$ & $7.213\times 10^{-2}$ & $8.448\times 10^{2}$ & -3.250 & -4.218  \\
40 & $8.866\times 10^{3}$ & 94.16 & $8.621\times 10^{-5}$ & $8.145\times 10^{-2}$ & $8.422\times 10^{2}$ & -3.218 & -4.630  \\
50 & $8.621\times 10^{3}$ & 92.85 & $8.929\times 10^{-5}$ & $8.704\times 10^{-2}$ & $9.285\times 10^{2}$ & -3.046 & -4.348  \\
60 & $1.229\times 10^{4}$ & 110.86 & $8.115\times 10^{-5}$ & $9.917\times 10^{-2}$ & $1.214\times 10^{3}$ & -3.200 & -4.840  \\
70 & $1.381\times 10^{4}$ & 117.54 & $8.355\times 10^{-5}$ & $1.083\times 10^{-1}$ & $1.391\times 10^{3}$ & -3.234 & -4.774  \\
80 & $1.418\times 10^{4}$ & 119.09 & $8.517\times 10^{-5}$ & $1.175\times 10^{-1}$ & $1.506\times 10^{3}$ & -3.290 & -4.760  \\
90 & $1.211\times 10^{4}$ & 110.05 & $8.621\times 10^{-5}$ & $1.246\times 10^{-1}$ & $1.476\times 10^{3}$ & -3.266 & -4.734  \\
100 & $1.383\times 10^{4}$ & 117.62 & $8.591\times 10^{-5}$ & $1.339\times 10^{-1}$ & $1.663\times 10^{3}$ & -3.352 & -4.822  \\
110 & $1.369\times 10^{4}$ & 117.01 & $8.563\times 10^{-5}$ & $1.399\times 10^{-1}$ & $1.736\times 10^{3}$ & -3.302 & -4.856  \\
120 & $1.426\times 10^{4}$ & 119.42 & $8.393\times 10^{-5}$ & $1.492\times 10^{-1}$ & $1.850\times 10^{3}$ & -3.340 & -5.078  \\
\bottomrule
\end{tabular}
\label{derived}
\end{table}

\begin{figure}[H]
    \centering
    \includegraphics[width=0.95\linewidth]{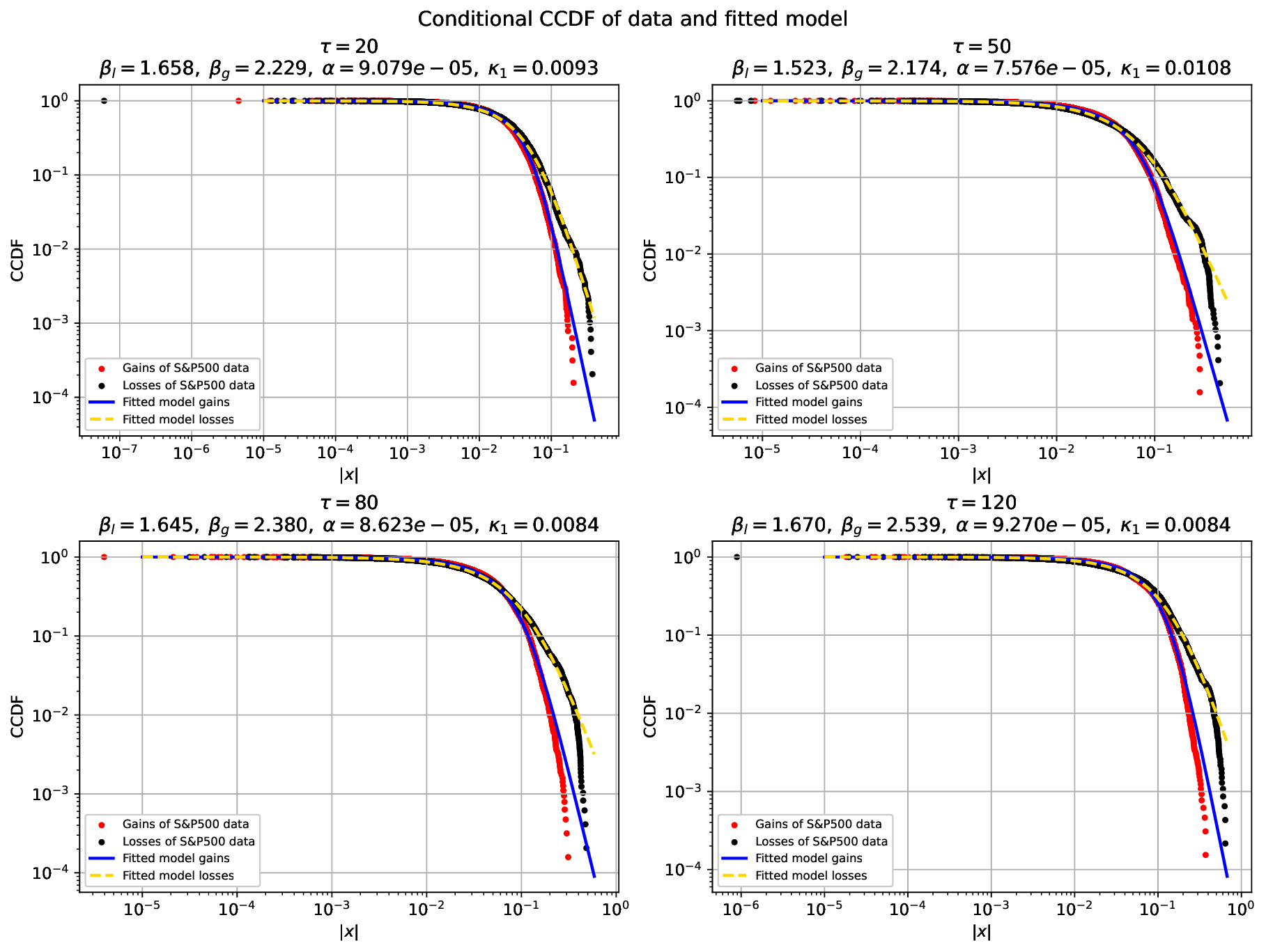}
    \caption{
    Representative CCDFs for accumulated returns. CCDF of S\&P500 gains and losses are plotted vis-a-vis those obtained from fitted distributions (\ref{ftmJFr}) with parameters from Table \ref{fitparams}.
    }
    \label{ccdf}
\end{figure}

CCDF of fitted distributions, with parameters from Table \ref{fitparams} (and Fig. \ref{fitparamsfig}), are shown next to CCDF of S\&P500 in Fig. \ref{ccdf}. It visually confirms slower tails for losses. Also the fidelity of fits to the actual S\&P500 data is quite good, including pre-bending areas of power-law tails. On the other hand, as obvious from Fig. \ref{tempering}, the actual bending point of fitted tempered mJF Skew-t happens at a point orders of magnitude larger than that of S\&P500 data.

\subsection{Statistical parameters\label{stats}}
\begin{figure}[H]
    \centering
    \includegraphics[width=0.63\linewidth]{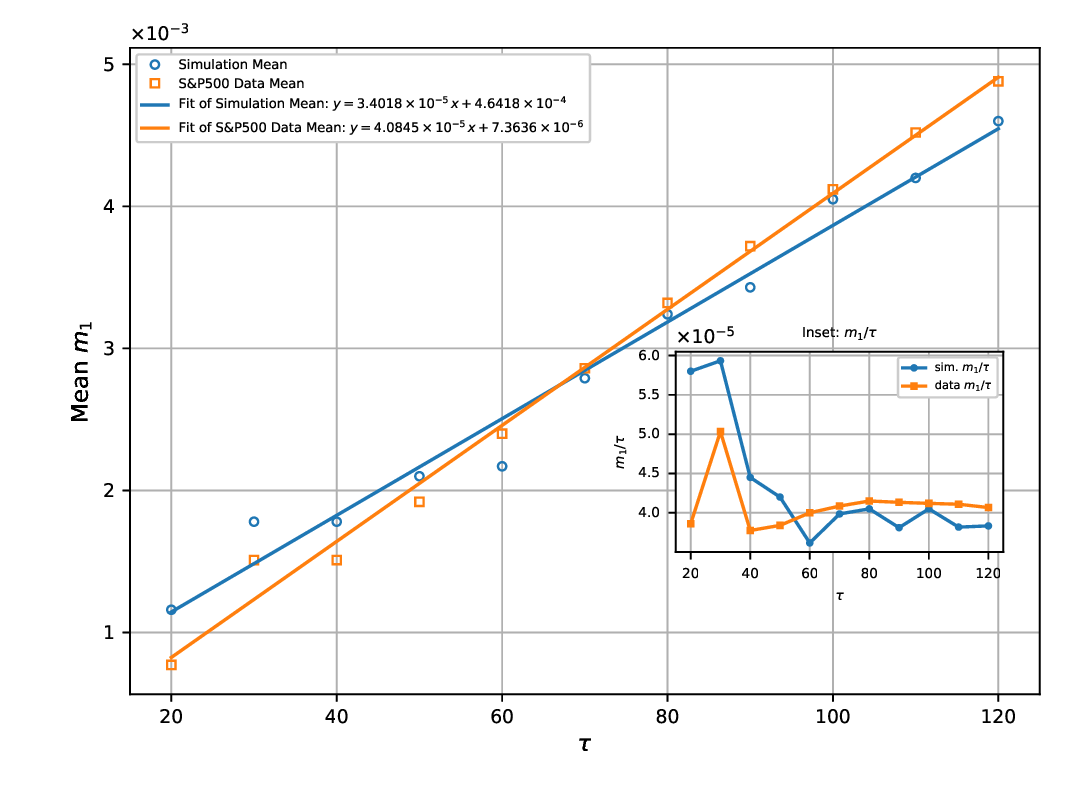}
    \caption{Mean of empirical S\&P500 data and of fitted distribution. Insert shows $m_1/\tau$.}
    \label{mean}
\end{figure}

\begin{figure}[H]
    \centering
    \includegraphics[width=0.63\linewidth]{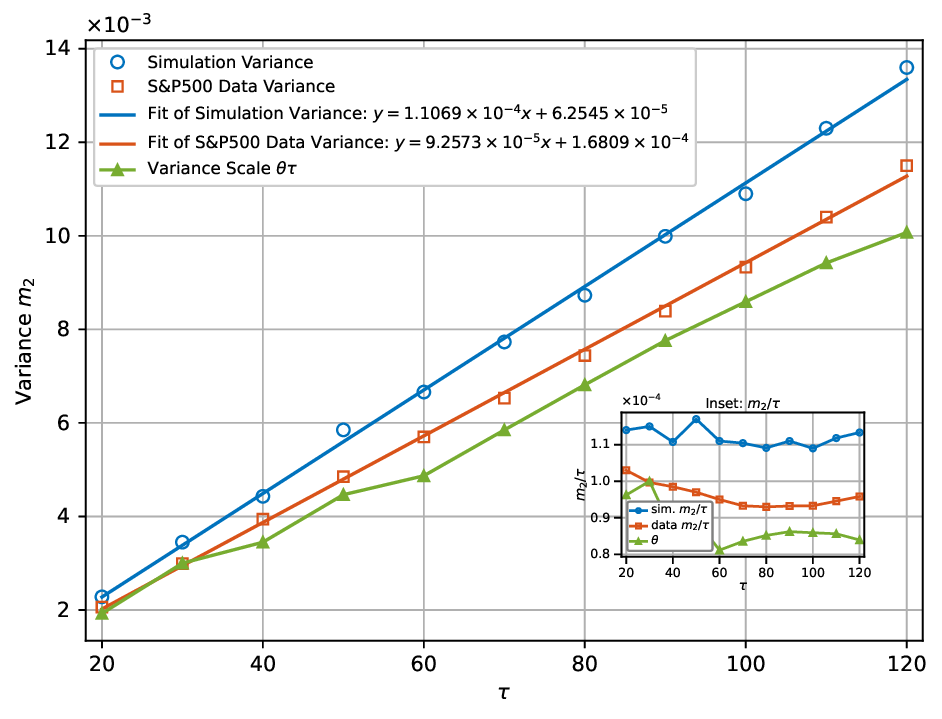}
    \caption{Variance of empirical S\&P500 data and of fitted distribution. Also shown $\theta\tau$ with $\theta$ obtained from (\ref{theta}). Insert shows $m_2/\tau$ and $\theta$.}
    \label{variance}
\end{figure}

Table \ref{statparams} contains values of mean $m_1$, variance $m_2$, median $\tilde{m}$, and mode $\bar{m}$ of S\&P500 data and those obtained by fitting with the tempered modified Jones-Faddy Skew-t distribution for returns (\ref{ftmJFr}) over $\tau=20,30,\ldots,120$ of days of accumulation of returns. The results contained in Table \ref{statparams} for mean $m_1$ and variance $m_2$ are graphically illustrated in Figs. \ref{mean} and \ref{variance}. 
Figure \ref{variance} also contains a plot of $\theta \tau$, where $\theta$ is defined by (\ref{theta}). It should be pointed out that for any mean-reverting model of stochastic variance, such as Cox-Ingersoll-Ross, \cite{cox1985theory,heston1993closed,dragulescu2002probability}, multiplicative model considered here (\ref{dvt}), and combined model \cite{dashti2021combined}, (\ref{dxt}) implies that
\begin{equation}
m_2=\langle\mathrm{d}x_t^2\rangle = \theta \tau
\label{thetatau}
\end{equation}
It is quite remarkable then that even when the symmetry is broken and power-law tails become tempered, variance $m_2$ still maintains linear dependence on $\tau$, as seen in Fig. \ref{variance}. We dubbed this phenomenon "conservation law" \cite{ghasemi2026broken}. We emphasize that, per discussion preceding (\ref{theta}), $\theta$ in Fig. \ref{variance} no longer directly related to mean reversion and is presented for guidance only. We also note that while some reasoning behind linear dependence of $m_2$ can be traced back to mean-reversion, linear dependence on $\tau$  of mean $m_1$ observed in Fig. \ref{mean} is not presently amenable to satisfactory explanation. 
\begin{table}[H]
\centering
\renewcommand{\arraystretch}{1.05}
\caption{Comparison of mean $m_1$, variance $m_2$, median $\tilde m$, and mode $\bar m$ between S\&P500 data and fitted distribution (\ref{ftmJFr}).}
\begin{tabular}{@{}c c c c c c c c c@{}}
\toprule
$\tau$ & $m_1$ data & $m_1$ sim. & $m_2$ data & $m_2$ sim. & $\tilde m$ data & $\tilde m$ sim. & $\bar m$ data & $\bar m$ sim.  \\
\midrule
20  & 0.000772 & 0.00116 & 0.00206 & 0.00228 & 0.00578 & 0.00498 & 0.0124 & 0.00856  \\
30  & 0.00151  & 0.00178 & 0.00299 & 0.00345 & 0.00635 & 0.00642 & 0.0193 & 0.0103  \\
40  & 0.00151  & 0.00178 & 0.00394 & 0.00443 & 0.00763 & 0.00819 & 0.00953 & 0.0142  \\
50  & 0.00192  & 0.00210 & 0.00485 & 0.00585 & 0.00893 & 0.00952 & 0.0118 & 0.0161  \\
60  & 0.00240  & 0.00217 & 0.00570 & 0.00666 & 0.0105  & 0.0109  & 0.0129 & 0.0191  \\
70  & 0.00286  & 0.00279 & 0.00653 & 0.00773 & 0.0115  & 0.0117  & 0.0212 & 0.0201  \\
80  & 0.00332  & 0.00324 & 0.00744 & 0.00873 & 0.0117  & 0.0123  & 0.0203 & 0.0208  \\
90  & 0.00372  & 0.00343 & 0.00839 & 0.00999 & 0.0116  & 0.0131  & 0.0155 & 0.0223  \\
100 & 0.00412  & 0.00405 & 0.00933 & 0.0109  & 0.0115  & 0.0139  & 0.0116 & 0.0233  \\
110 & 0.00452  & 0.00420 & 0.0104  & 0.0123  & 0.0138  & 0.0153  & 0.0137 & 0.0258  \\
120 & 0.00488  & 0.00460 & 0.0115  & 0.0136  & 0.0157  & 0.0169  & 0.0206 & 0.0288  \\
\bottomrule
\end{tabular}
\label{statparams}
\end{table}

\subsection{Skewness\label{skew}}
\begin{figure}[H]
    \centering
    \includegraphics[width=0.49\linewidth]{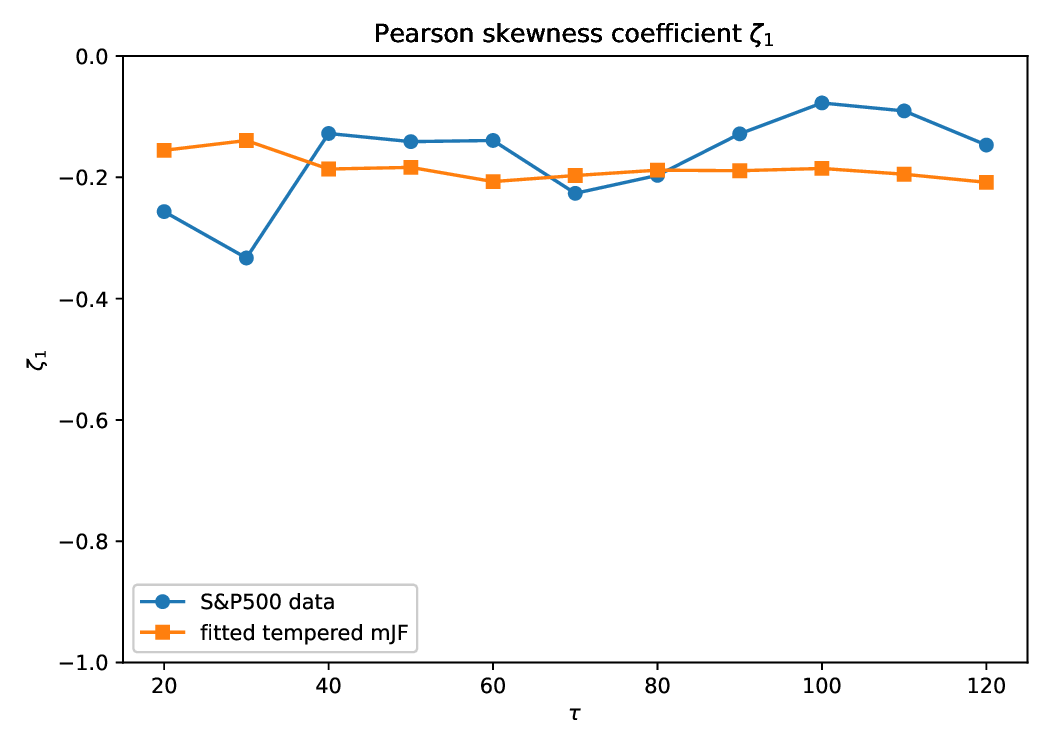}
    \caption{First Pearson skewness coefficients for S\&P500 data and fitted distribution as a function of number of days of accumulation of returns.}
    \label{skewplot1}
\end{figure}

\begin{figure}[H]
    \centering
    \includegraphics[width=0.49\linewidth]{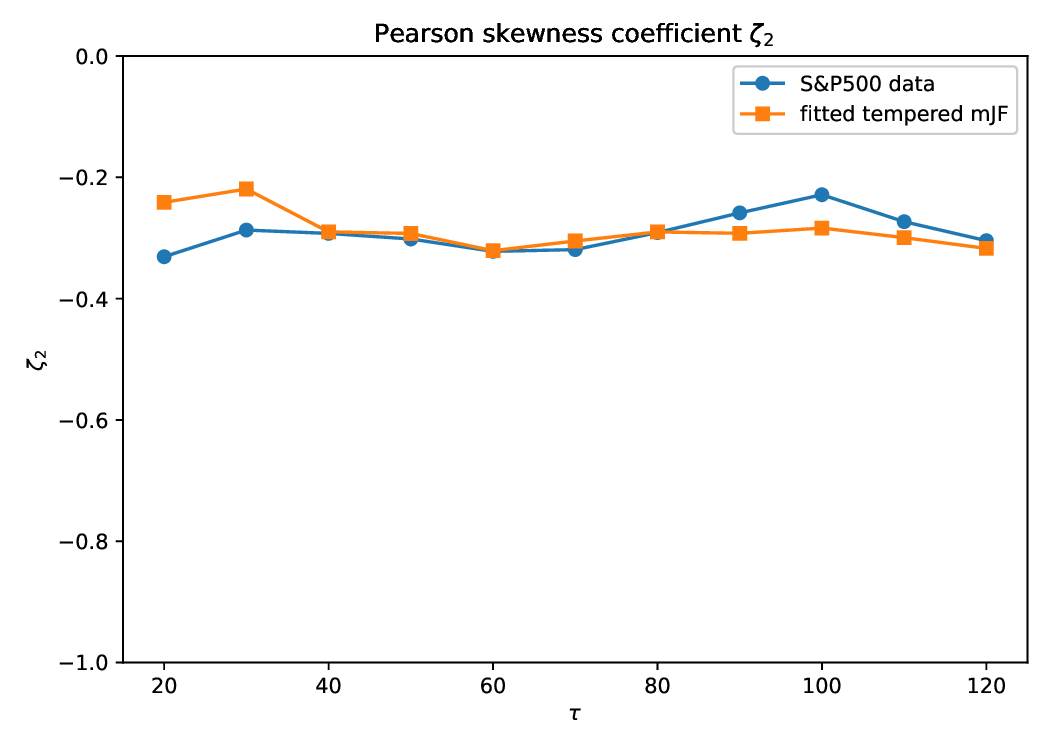}
    \caption{Second Pearson skewness coefficients for S\&P500 data and fitted distribution as a function of number of days of accumulation of returns.}
    \label{skewplot2}
\end{figure}

\begin{figure}[H]
    \centering
    \includegraphics[width=0.49\linewidth]{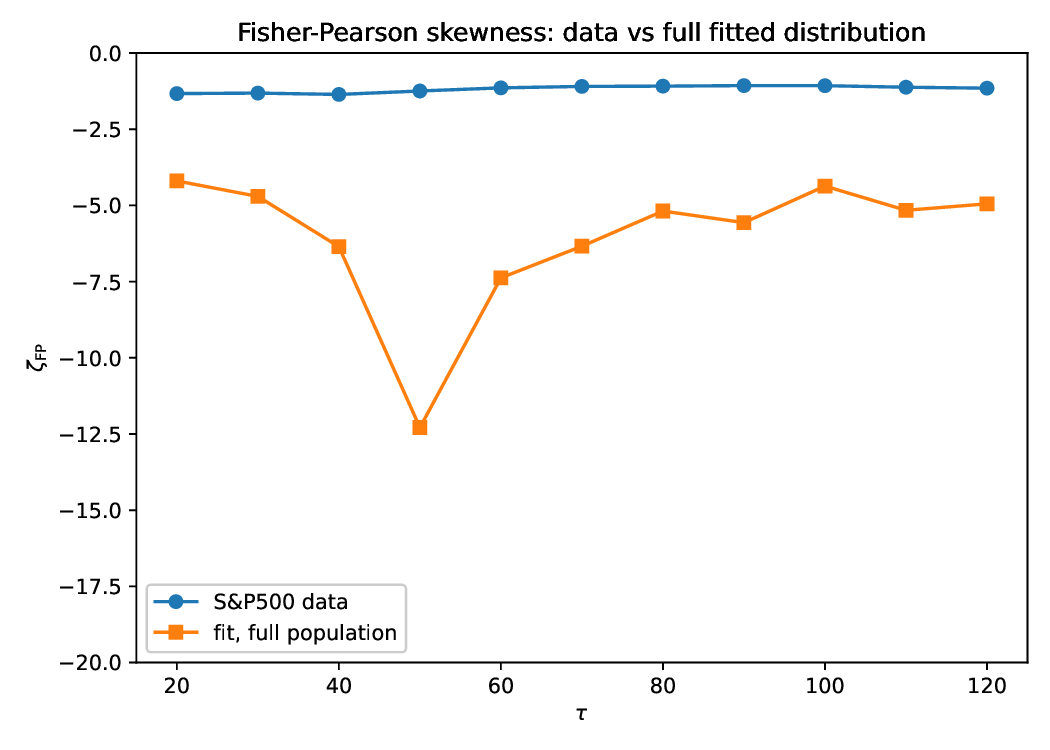}
  \caption{Fisher--Pearson skewness coefficient for S\&P500 data and for the full fitted tempered mJF Skew-t distribution as a function of the number of days of accumulation of returns.}
    \label{skewplot3}
\end{figure}

To characterize skewness, we use the first and second Pearson coefficients \cite{pearson1984mathematical,pearson1985mathematical}, as well as the Fisher--Pearson coefficient \cite{fisher1930moments} 
\begin{equation}
 \zeta_1=\frac{m_1-\bar m}{\sqrt{m_2}},
 \qquad
 \zeta_2=\frac{3(m_1-\tilde m)}{\sqrt{m_2}},
  \qquad
 \zeta_{\rm FP}=\frac{m_3}{m_2^{3/2}},
 \label{eq:zeta}
\end{equation}
where $m_1$ is the mean, $m_2$ is the variance, $\bar m$ is the mode, $\tilde m$ is the median, and $m_3=\langle (x-m_1)^3\rangle$ is the third central moment.  In the previous pure-power-law setting \cite{ghasemi2026broken}, the Fisher--Pearson coefficient may fail to exist for sufficiently heavy tails.  In the present tempered distributions, however, the tails are regularized, so the third central moment is finite and $\zeta_{\rm FP}$ may in principle serve as meaningful additional diagnostic. Table \ref{skewness} shows $\zeta_1$, $\zeta_2$ and $\zeta_{\rm FP}$ for S\&P500 data and those obtained from fitted distributions for $\tau=20,30,\ldots,120$ days of accumulations of returns. Graphical representation of the Pearson coefficients in Table \ref{skewness} is given in Figs.~\ref{skewplot1}--\ref{skewplot3}.
\begin{table}[H]
\centering
\caption{First and second Pearson skewness coefficients, $\zeta_1$ and $\zeta_2$, and Fisher--Pearson skewness coefficient $\zeta_{\rm FP}$ for S\&P500 data and fitted tempered distributions.}
\small
\begin{tabular}{@{}c c c c c c c@{}}
\toprule
$\tau$ & $\zeta_1$ data & $\zeta_1$ sim. & $\zeta_2$ data & $\zeta_2$ sim. & $\zeta_{\rm FP}$ data & $\zeta_{\rm FP}$ sim. \\
\midrule
20  & -0.2565 & -0.1551 & -0.3310 & -0.2400 & -1.3316 & -4.1973  \\
30  & -0.3259 & -0.1452 & -0.2658 & -0.2367 & -1.3161 & -4.7057  \\
40  & -0.1277 & -0.1860 & -0.2925 & -0.2888 & -1.3598 & -6.3547  \\
50  & -0.1412 & -0.1831 & -0.3017 & -0.2910 & -1.2468 & -12.2880 \\
60  & -0.1392 & -0.2068 & -0.3221 & -0.3201 & -1.1438 & -7.3774  \\
70  & -0.2264 & -0.1968 & -0.3192 & -0.3043 & -1.0967 & -6.3386  \\
80  & -0.1968 & -0.1881 & -0.2911 & -0.2897 & -1.0875 & -5.1850  \\
90  & -0.1283 & -0.1889 & -0.2587 & -0.2916 & -1.0694 & -5.5665  \\
100 & -0.0774 & -0.1850 & -0.2287 & -0.2830 & -1.0725 & -4.3661  \\
110 & -0.0906 & -0.1950 & -0.2732 & -0.2998 & -1.1226 & -5.1608  \\
120 & -0.1468 & -0.2080 & -0.3044 & -0.3165 & -1.1541 & -4.9484  \\
\bottomrule
\end{tabular}
\label{skewness}
\end{table}
Clearly, fidelity between the data and the simulation is quite good for $\zeta_1$ and $\zeta_2$. Also, skewness depends little on the number of the days of accumulation and its negative sign is in strong agreement with the heavier tails of the losses. On the other hand, discrepancy in $\zeta_{FP}$ between data and simulation is completely understandable given the fact, pointed out in Sec. \ref{params}, that power-law tails of fitted tempered mJF Skew-t distributions continue for orders of magnitude longer than those of S\&P500.

\section {Scaling \label{scaling}}
Scaling has been long recognized as a feature of fluctuations in asset prices \cite{mantegna1995scaling,gopikrishnan1999scaling,fuentes2009universal,ghasemi2026broken}. Tempered mJF Skew-t distribution for returns fits this mold as well. Indeed, introducing a new variable
\[
y=\frac{x-\mu}{\sqrt{2 \alpha\tau}}
\]
and introducing a renormalized tempering parameter 
\[
\tilde{\kappa}_1=\kappa_1 \alpha^{1/2}
\]
(\ref{ftmJFr}) can be rewritten in a rescaled form as
\begin{align}
\tilde{f}_{\rm tJF} (y)= \frac{1}{\tilde{Z}}\Bigg\{&
\frac{3\tilde{\kappa}_1^{1+2\bar\beta}
\Gamma\!\left(-\frac12-\bar\beta\right)\Gamma(\bar\beta-1)}
{4\sqrt{\pi}}\,
{}_1F_1\!\left(
\frac52;\frac32+\bar\beta;
-\tilde{\kappa}_1^2 \left(y^2+1\right)
\right)
\notag\\[3pt]
&\quad
+
\Gamma\!\left(\bar\beta+\frac12\right)
\left(
1+\frac{y}{\sqrt{y^2+1}}
\right)^{\beta_l+\frac12}
\left[
1-\frac{y}{\sqrt{y^2+1}}
\right]^{\beta_g+\frac12}
\notag\\[3pt]
&\qquad\qquad\times
{}_1F_1\!\left(
2-\bar\beta;\frac12-\bar\beta;
-\tilde{\kappa}_1^2 \left(y^2+1\right)
\right)\Bigg\}, \qquad \bar{\beta}=\frac{\beta_l+\beta_g}{2}
\label{ftJFscaled}
\end{align}
thus eliminating dependence on $\tau$. In this form, PDF (\ref{ftJFscaled}) can be considered as a generic form of tempered Jones-Faddy Skew-t distribution, now divorced from the specific application to asset returns. In particular, shape parameters $\beta_{l,g}$, previously for "losses" and "gains," should now be understood as $\beta_{l,r}$ for left and right tails.

It should be pointed out that via change of variable
\[
y=\frac{x}{\sqrt{\bar{\nu}}}
\]
PDF of generic Jones-Faddy Skew-t distribution (\ref{fJF}) can also be written in a rescaled form 
\begin{equation}
\tilde{f}_{\rm JF}(y)= \frac{1}{2^{\bar{\nu} -1} B\left(\frac{\nu_l}{2}, \frac{\nu_r}{2} \right)} \left(1+\frac{y}{\sqrt{y^2+1}}\right)^{\frac{\nu_l+1}{2}} \left(1-\frac{y}{\sqrt{y^2+1}}\right)^{\frac{\nu_r+1}{2}}, \qquad \bar{\nu} = \frac{\nu_l + \nu_r}{2}
\label{fJFscaled}
\end{equation}
which elucidates the fact that tempered (\ref{ftJFscaled}) is the generalized version of (\ref{fJFscaled}) via tempering parameter $\tilde{\kappa}_1$. Obviously, setting $\beta_l=\beta_g=\beta$ in (\ref{ftJFscaled}) and $\nu_l=\nu_r=\nu$ in (\ref{fJFscaled}) would render scaled versions of tempered Student t- and Student t- (\ref{fst}) distributions respectively.

\begin{figure}[H]
    \centering
    \includegraphics[width=0.72\linewidth]{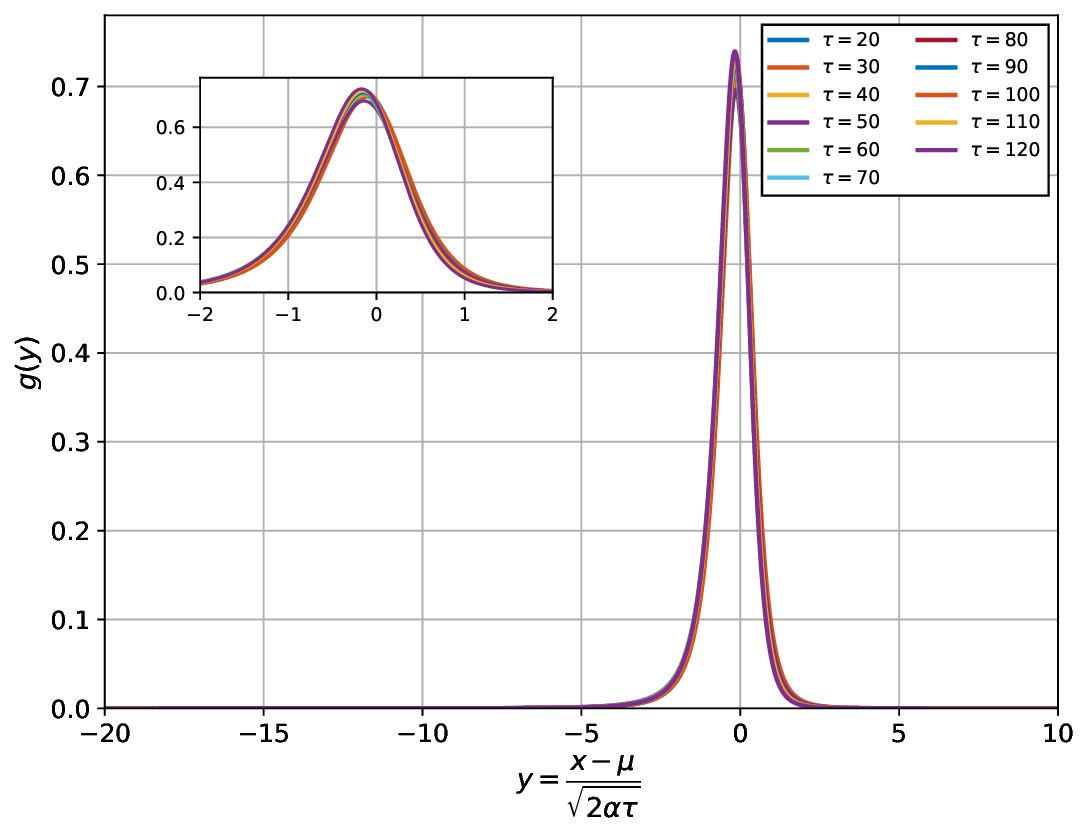}
    \caption{PDF (\ref{ftmJFr}) for $\tau=20,30,\ldots,120$ as a function of $(x-\mu)/\sqrt{2 \alpha \tau}$ using parameters from Table \ref{fitparams}.}
    \label{pdfscaled}
\end{figure}

\begin{figure}[H]
    \centering
    \begin{subfigure}{0.48\linewidth}
    \centering
    \includegraphics[width=\linewidth]{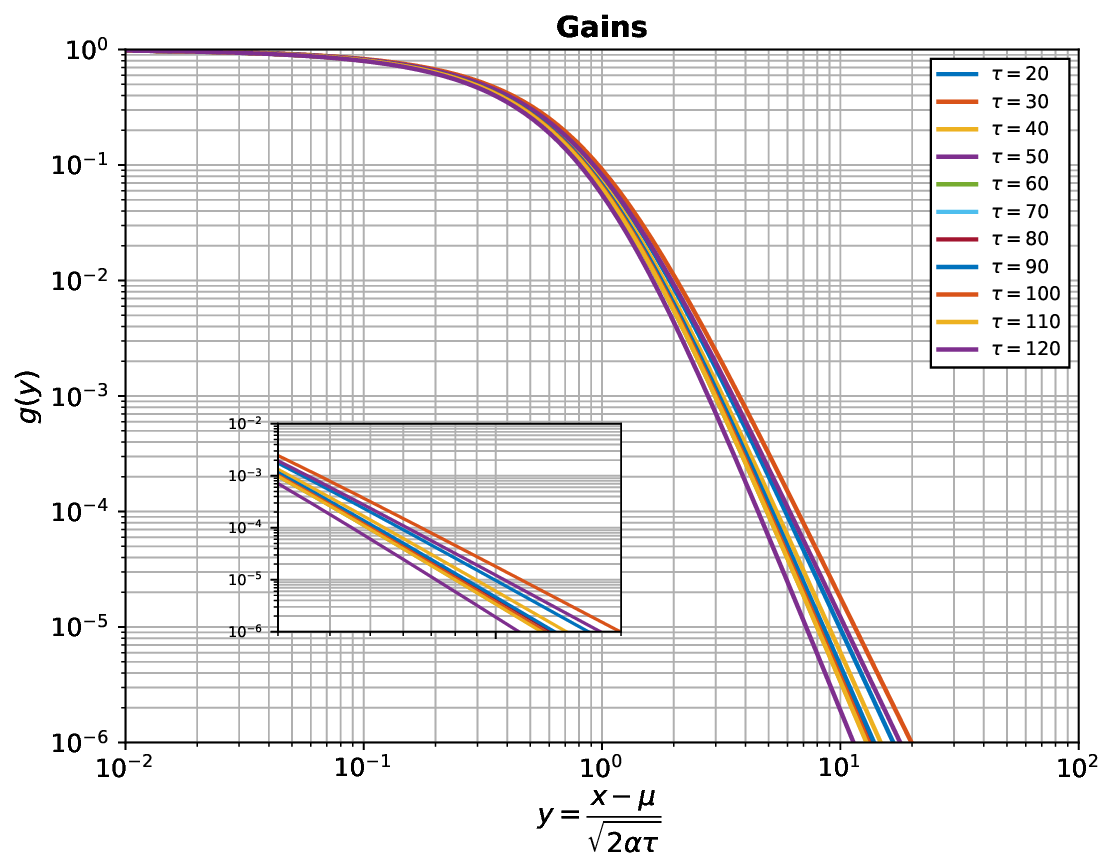}
    \caption{Gains.}
    \end{subfigure}
    \hfill
    \begin{subfigure}{0.48\linewidth}
    \centering
    \includegraphics[width=\linewidth]{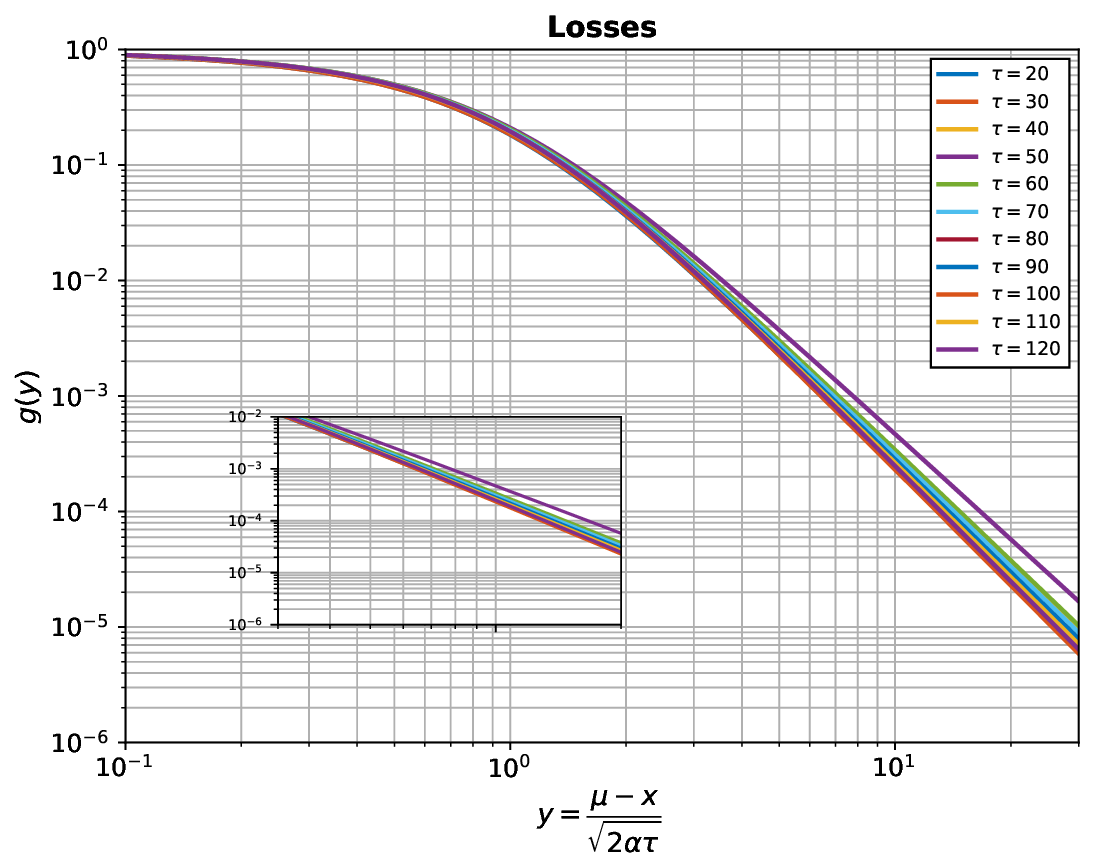}
    \caption{Losses.}
    \end{subfigure}
    \caption{CCDF, corresponding to PDF (\ref{ftmJFr}), for $\tau=20,30,\ldots,120$ as a function of $(x-\mu)/\sqrt{2 \alpha \tau}$ using parameters from Table \ref{fitparams}. Inserts show tail areas of CCDF.}
    \label{ccdfscaled}
\end{figure}

Figs. \ref{pdfscaled} and \ref{ccdfscaled} show PDF (\ref{ftmJFr}) and the corresponding CCDF for $\tau=20,30,\ldots,120$ as a function of $(x-\mu)/\sqrt{2 \alpha \tau}$ using parameters from Table \ref{fitparams}. Clearly, these figures comport well with the idea of scaling embodied in (\ref{ftJFscaled}).

\clearpage
\section{Discussion \label{discuss}}
The main result of this work is the introduction of three new distributions. First, we introduced IGa1 distribution whose PDF is given by (\ref{tfvt}) and (\ref{tfvtparams}). This distribution is capped at finite value $\kappa_1^{-2}$.  However, if $\alpha \ll  \kappa_1^{-2}$ then IGa1 would exhibit standard IGa power-law tails $\propto v_t^{-\beta-1}$ for $\alpha \ll v_t \ll \kappa_1^{-2}$. Second, we introduced a tempered Student t-distribution, (\ref{ftStr}) and (\ref{Ix}), which is obtained as a product distribution of IGa1 with Gaussian increments of the Wiener process and, as such, exhibits power-law tails with the same exponents as IGa1 between points $x_o$ and $x_b$ estimated in (\ref{onset}) and (\ref{bend}). While this distribution is not capped due to the normal distribution's infinite support interval, it exhibits significant downward bending after $x_b$, as seen in Fig. \ref{tempering}. Lastly, using Jones-Faddy procedure, we broke symmetry of the tempered Student t-distribution to obtain a tempered modified Jones-Faddy Skew-t distribution, (\ref{ftmJFr}) and (\ref{Ixgl}), which is characterized by unequal left and right power-law exponents prior to tempering.

We applied tempered modified Jones-Faddy Skew-t distribution to analysis of multi-day S\&P500 returns for the number of days of accumulation from 20 to 120. Remarkably, the variance of returns (realized variance) shows consistent linear dependence on the number of days of accumulation both for S\&P500 data and for fitted distributions, as seen in Fig. \ref{variance}. While it is expected for symmetrical distributions according to (\ref{dxt}), it is not at all obvious why this should be the case for asymmetrical, tempered distributions as is the case for empirical data. Furthermore, linear dependence on the number of days of accumulation is also observed for the mean of both empirical and fitted distributions, as seen in Fig. \ref{mean}. We hope to address the underlying reason for the latter in a future work.

Goodness of fit with tempered modified Jones-Faddy Skew-t distribution is also confirmed by comparison of the coefficients of skewness in Figs. \ref{skewplot1} and \ref{skewplot2}. On the other hand, tempering scale predicted by this distribution is orders of magnitude off from that observed in empirical data. This most likely indicates that goodness of fit is little affected by the few points in tempered end-tails. However, the discrepancy between tempering scales of empirical data and the fits is responsible for misalignment in Fig. \ref{skewplot3}. Finally, in Figs. \ref{pdfscaled} and \ref{ccdfscaled} we observe good compliance with scaling prediction in accordance with (\ref{ftJFscaled}). While we were specifically addressing asset returns, (\ref{ftJFscaled}) can be considered as a generic rescaled tempered Jones-Faddy distribution. The same consideration can be easily applied to rescaled versions of IGa1 and tempered Student t-distribution. 

Aside from the above-mentioned issues left for future work, we also hope to investigate other indices and large-cap individual companies, both domestic and international, to see how prevalent -- or not -- our findings for S\&P500 are. Additionally, it will be interesting to see whether tempered Jones-Faddy distribution can be applied to other asset classes, such as housing prices. Finally, we are designing more robust tools for testing goodness of fit, which is particularly important for distributions with power-law tails.

\clearpage

\bibliography{mybib}

\end{document}